\documentclass[]{aastex631}

\usepackage{newtxtext,newtxmath}
\usepackage[T1]{fontenc}
\usepackage[encapsulated]{CJK}

\begin{document}
\begin{CJK*}{UTF8}{gbsn}
\title{Study of Advective Energy Transport in the Inflow and the Outflow of Super-Eddington Accretion Flows}

\author[0000-0002-7663-7900]{Cheng-Liang Jiao (焦承亮)}
\affiliation{Yunnan Observatories, Chinese Academy of Sciences, 396 Yangfangwang, Guandu District, Kunming, 650216, P. R. China}
\affiliation{Center for Astronomical Mega-Science, Chinese Academy of Sciences, 20A Datun Road, Chaoyang District, Beijing, 100012, P. R. China}
\affiliation{Key Laboratory for the Structure and Evolution of Celestial Objects, Chinese Academy of Sciences, 396 Yangfangwang, Guandu District, Kunming, 650216, P. R. China} 
\correspondingauthor{Cheng-Liang Jiao (焦承亮)}
\email{jiaocl@ynao.ac.cn}

\begin{abstract}
	Photon trapping is believed to be an important mechanism in super-Eddington accretion, which greatly reduces the radiative efficiency as photons are swallowed by the central black hole before they can escape from the accretion flow. This effect is interpreted as the radial advection of energy in one-dimensional height-integrated models, such as the slim disk model. However, when multi-dimensional effects are considered, the conventional understanding may no longer hold. In this paper, we study the advective energy transport in super-Eddington accretion, based on a new two-dimensional inflow-outflow solution with radial self-similarity, in which the advective factor is calculated self-consistently by incorporating the calculation of radiative flux, instead of being set as an input parameter. We found that radial advection is actually a heating mechanism in the inflow due to compression, and the energy balance in the inflow is maintained by cooling via radiation and vertical ($\theta$-direction) advection, which transports entropy upwards to be radiated closer to the surface or carried away by the outflow. As a result, less photons are advected inwards and more photons are released from the surface, so that the mean advective factor is smaller and the emergent flux is larger than those predicted by the slim disk model. The radiative efficiency of super-Eddington accretion thus should be larger than that of the slim disk model, which agrees with the results of some recent numerical simulations.
\end{abstract}

\section{INTRODUCTION}\label{intro}
Accretion flows are called "super-Eddington" when mass accretion rates are greater than the Eddington value, $\dot{M}_\mathrm{Edd} \equiv L_\mathrm{Edd}/(\eta c^2)$, where $L_\mathrm{Edd}$ ($=4 \pi c G M/\kappa_\mathrm{es}$) is the Eddington luminosity and $\eta$ is the radiative efficiency. The high density, and consequently the large optical depth, delays the liberation of radiation energy, so that part of the energy generated by local viscous dissipation is not radiated away and instead trapped in the flow and advected inward, which is called photon trapping \citep[e.g.,][]{Katz1977, Begelman1978, Ohsuga2002}.
While its general reference in literature is connected with trapped photons swallowed by the central black hole (BH) and the consequent reduction in radiative efficiency, photon trapping actually has two different interpretations when quantified in models.
The first interpretation is fairly straight-forward and focuses on the accreted matter, so that photon trapping is quantified by $E_\mathrm{rad}\boldsymbol{v}$ \citep[e.g.,][]{Ohsuga2005, Ohsuga2009, OM2007, OM2011}, where $E_\mathrm{rad}$ is the radiation energy density. In this case, photon trapping is just a phenomenon, a natural result of imbalance in the energy equation, where radiation is not enough to cool the heating caused by viscous dissipation so that the left-over energy is stored in the radiation field (photons) coupled with matter and carried around.
The second interpretation is widely adopted in steady models of super-Eddington accretion, e.g. the slim disk model \citep{Ab88,CT93,Jiao09,Sadowski2011}, and focuses on a fixed region in the rest frame, so that photon trapping is quantified by $\boldsymbol{\nabla \boldsymbol{\cdot}} (E_\mathrm{rad}\boldsymbol{v})$. In this case, as matter is accreted towards the central BH, it is expected that more and more photons are trapped, so that the outgoing amount of radiation energy for the fixed region (caused by motion instead of radiative flux)  is larger than the incoming amount, and the net loss in radiation energy effectively cools the region, which participates in the energy balance in steady models and is part of the advective energy transport (usually referred to as "advective cooling" in conventional steady models), which represents the net change of total entropy in the fixed region. 
The "advective cooling" scenario is confirmed in one-dimensional height-integrated models such as the slim disk model, where the vertical variation of radial velocity $v_r$ and the vertical motion are neglected\footnote{The vertical dependences of other physical quantities are also assumed in the slim disk model (usually one of the the one-zone, isothermal, or polytropic assumptions), which also affects the radiative efficiency \citep{Kohri2007}.}. However, these simplifications in vertical direction originally arise from early studies of accretion disks which are assumed to be very thin [e.g., the standard thin disk (SSD); \citealp{SS73,NT73}], and are not necessarily applicable to super-Eddington accretion disks whose height is at least comparable to radius. If we consider a more realistic vertical structure of the accretion flow and allows multi-dimensional effects, then this interpretation of photon trapping becomes problematic.

The first concern is the existence of outflow, i.e., the region with positive radial velocity where matter escapes from the central BH. While outflow is neglected in the slim disk model due to its simplification in vertical direction, recent studies suggest that it does exist in super-Eddington accretion flows. Theoretically speaking, outflow will be inevitably developed when the accretion rate becomes large enough, as the radiative force will eventually overcome the gravitational force to drive the outflow \citep{HD07,GL07,Jiao09}. Outflow is also commonly found in numerical simulations of super-Eddington accretion \citep[e.g.,][]{Ohsuga2005, Ohsuga2009, OM2007, OM2011, Yang2014, Sadowski2014, Sadowski2015, McKinney2014, Jiang2014, Jiang2019, Kitaki2018, Kitaki2021, Asahina2022}, though the strength of outflow is still debatable. Observationally, recent discoveries suggest that most of the ultraluminous X-ray sources (ULXs) appear to be super-Eddington accretors with powerful outflows \citep{Fabrika2021,King2023}. The radial motion in the inflow and the outflow are opposite, which naturally means a different sign in the advective energy transport (when vertical motion is neglected). So if the advective energy transport is a cooling mechanism in one region, then it must be heating in the other region. Note that in the conventional slim disk model, the "advective cooling" is an important (or even dominant for very high accretion rates) cooling mechanism, so how the energy equation is balanced in both the inflow and the outflow would become problematic. The second concern is that if vertical motion is allowed, it can possibly carry photons upward towards the surface, where photons can more easily escape due to the reduced optical depth. This effect can be described by the vertical advection. \cite{Ohsuga2002} already mentioned the possibility that photon trapping may be attenuated in the presence of large-scale circulation motion, which could considerably reduce the time for photons travelling to the surface. This effect is verified in their later numerical simulation \citep{OM2007}. The recent simulations of super-Eddington accretion performed by \cite{Jiang2014,Jiang2019} investigated this effect in more detail, and found that vertical advection of radiation field allows a significant fraction of photons to escape before being advected into the BH, so that the radiative efficiency is much larger than that predicted by the slim disk model. Still, these effects remain to be studied in steady models for super-Eddington accretion.

In this paper, we investigate these effects in steady models. To do this, we need to obtain detailed structure of the accretion flow in the vertical direction. Radial self-similarity under axisymmetry is a powerful tool, enabling us to reduce the partial differential equations (PDEs) describing the accretion flow into ordinary differential equations (ODEs) in $\theta$ direction in spherical coordinates ($r\theta\phi$), which can then be solved with proper boundary conditions. We have studied the advective energy transport in the inflow and the outflow of optically thin advection-dominated accretion flows \citep[ADAFs;][]{Jiao2022} based on the two-dimensional (2D) self-similar model established by \cite{JW11}. However, that model is not applicable to the study in this paper. The reason is that the advective factor $f$, which represents the ratio of the advective energy transport to the viscous heating, is set to be a constant input parameter in \cite{JW11} and \cite{Jiao2022}. While this can be justified in typical ADAFs due to their radiative inefficiency (so that $f \to 1$), it is not so well defined in super-Eddington accretion flows whose radiation is non-negligible. 
As for other related works in literature, \cite{Gu2012} calculated the vertical structure of radiation pressure-supported accretion disks, and \cite{Samadi2019} calculated the vertical structure of accretion disks with comparable gas and radiation pressure, both with detailed calculations of radiation and consequently a variable $f$, but they both set $v_\theta=0$ so that there is neither outflow nor vertical advection in their solutions.
\cite{Zeraatgari2020} presented 2D inflow-outflow solutions of super-Eddington accretion flows with radial density index $n=1/2$ (for $\rho \propto r^{-n}$) and a variable $f$. The problem is that they applied the symmetric boundary conditions on both the equatorial plane and the polar axis, in which case the integration of the continuity equation along $\theta$ direction would yield a net accretion rate of 0 for $n \neq 3/2$ \cite[see][for a detailed derivation]{Jiao2022}. While this may happen temporarily for an ADAF with very low accretion rate, it is hardly applicable to super-Eddington accretion flows, and the super-Eddington luminosities observed in ULXs \citep{Fabrika2021,King2023} definitely require large net accretion rates of matter whose gravitational energy powers the radiation eventually.
So we have to formulate a new 2D self-similar model with a variable advective factor $f$ in this paper. We focus on the region with typical features of super-Eddington accretion, such as radiation-pressure domination and the domination of electron scattering in opacity. The energy transport via radiation and advection are calculated in detail, which together balances the heating of viscous dissipation. The advective energy transport in the inflow and the outflow is then analysed in detail based on mechanism (advected internal energy and compression heating/expansion cooling) and direction ($r$ and $\theta$ directions). 
The influence of vertical advection on the radiative efficiency is also investigated.

The paper is arranged as follows. The new model is presented in Section \ref{model}. The numerical results are presented and discussed in Section \ref{result}. A summary is presented in Section \ref{summary}.

\section{THE MODEL}\label{model}
\subsection{Basic Equations and Assumptions}\label{eqs}
We consider a steady($\partial /\partial t=0$) and axisymmetric($\partial
/\partial \phi=0$) accretion flow onto a non-spinning BH at the origin of spherical coordinates ($r\theta\phi$), neglecting self-gravity of accreted matter. The equations of continuity and motion can be respectively written as
\begin{equation}\label{eq_continuity}
	{\nabla}\boldsymbol{\cdot}(\rho \boldsymbol{v})=0,
\end{equation}
\begin{equation}\label{eq_motion}
	\rho (\boldsymbol{v} \boldsymbol{\cdot} {\nabla}) \boldsymbol{v}= -\rho {\nabla} \Psi -{\nabla} p_\mathrm{gas} + {\nabla} \boldsymbol{\cdot} \textsf{\textbf{T}} + \rho \frac{\kappa}{c}\boldsymbol{F},
\end{equation}
where $\Psi$ is the gravitational potential, $\textsf{\textbf{T}}$ the tensor of viscous stress, $\boldsymbol{F}$ the radiative flux, and $\kappa$ the total opacity. We assume that for the accretion flow, the $r\phi$-component of the viscous stress tensor, $t_{r\phi}$, is dominant, and adopt the $\alpha$ prescription of viscosity \citep{SS73}, $t_{r\phi}=-\alpha p$, where $p=p_\mathrm{gas}+p_\mathrm{rad}$ is the total pressure. We adopt the Newtonian gravitational potential, $\Psi=-GM/r$, in this paper. Note that close to the central BH, relativistic effects become strong and self-similarity may no longer hold, so we keep our study beyond that region and focus on the region in the range of intermediate radii away from the inner and the outer boundaries, where self-similarity is a good approximation \citep{NY94,Narayan97,Jiao15}. Thus Newtonian potential is good enough in our calculation.

For a typical super-Eddington accretion flow, both steady models and numerical simulations have shown that the principal part of the flow should be both optically thick and extremely radiation-pressure dominated (the ratio of gas pressure to total pressure  $\beta \to 0$), ranging from the inner edge of the flow to several hundred or thousand times of Schwarzschild radius $r_\mathrm{S}$ $(\equiv 2GM/c^2)$, depending on the accretion rate. The accretion flow beyond this range generally resembles the structure of a SSD, though it should still be optically thick and radiation-pressure dominated, up to a much larger radius \citep[][p.112]{Kato2008}. 
Here we focus the study on the typical region of a super-Eddington accretion flow. 
As the flow is optically thick, we can adopt the Eddington approximation, $p_{\mathrm{rad}}= E_{\mathrm{rad}}/3$, where $E_{\mathrm{rad}}$ is the radiation energy density (per unit volume). According to the first moment of the radiative transfer equation, for a steady radiation field, we have \citep[][p.490]{Kato2008} 
\begin{equation}\label{eq_flux}
	\boldsymbol{F}=-\frac{c}{\rho \kappa} {\boldsymbol{\nabla}} p_\mathrm{rad},
\end{equation}
so in Equation (\ref{eq_motion}) the radiation force term can actually be incorporated into the gas-pressure gradient term to form the gradient of the total pressure, $p=p_\mathrm{gas}+p_\mathrm{rad}$, which is the form of the equation of motion adopted in \cite{XW05} and \cite{JW11}.

The energy equation can be written as (\citealp{Ohsuga2005}; also see \citealp{Jiao15}),
\begin{equation}\label{eq_energy}
	\boldsymbol{\nabla \boldsymbol{\cdot}} (E \boldsymbol{v}) + p \boldsymbol{\nabla \boldsymbol{\cdot} v} = q_\mathrm{vis} - \boldsymbol{\nabla \boldsymbol{\cdot} F},
\end{equation} 
where $E=E_\mathrm{gas}+E_\mathrm{rad}$ is the total energy density of the coupled matter and radiation. As mentioned above, gas pressure can be neglected in a typical super-Eddington accretion flow, so here we actually use $p=p_\mathrm{rad}$ and $E=E_\mathrm{rad}=3p$, and the left hand side (LHS) of Equation (\ref{eq_energy}) essentially represents the advection of radiation field. On the right hand side (RHS), $q_\mathrm{vis}$ is the viscous heating rate, which can be expressed as \citep[][p.239]{Kato2008}
\begin{equation}\label{eq_qvis}
	q_\mathrm{vis}=t_{r\phi} r \frac{\partial }{\partial r}(\frac{v_\phi}{r}),
\end{equation}
and $\nabla \boldsymbol{\cdot} \boldsymbol{F}$ represents the change of the energy density in a fixed region due to the transportation via the radiative flux $\boldsymbol{F}$, which is usually referred to as "radiative cooling" term in height-integrated accretion disk models, though it could actually be either positive or negative here.

To close this set of equations, we need to know the opacity $\kappa$. We consider the low metallicity case in this paper. The scattering opacity comes from electron scattering, so we have $\kappa_\mathrm{sca} = \kappa_\mathrm{es} \approx 0.4$ cm$^2$ g$^{-1}$. The absorption opacity is dominated by the free-free absorption, $\kappa_\mathrm{ff} \propto \rho T^{-7/2}$. In super-Eddington accretion flows, temperature is relatively high, and $\kappa_{\mathrm{es}}$ generally dominates over $\kappa_{\mathrm{ff}}$, so here we can safely neglect the absorption opacity and set $\kappa=0.4$ cm$^2$ g$^{-1}$. Note that for the high metallicity case, the bound-free absorption opacity can play an important role \citep{Ohsuga2005}, especially in the outflow region, which could be the cause of blue-shifted absorption lines in some broad absorption line quasars. This will be studied in our future work.

With some mathematical deduction, Equations (\ref{eq_continuity}), (\ref{eq_motion}) and (\ref{eq_energy}) can be transformed to 5 PDEs of $r$ and $\theta$ with 5 unknowns, $\rho$, $v_r$, $v_\theta$, $v_\phi$, and $p$.
We adopt self-similar assumptions in the radial direction, so that the solution takes the form
\begin{eqnarray}
	\rho &=& \rho(\theta)r^{-n}, \label{eq_ss1} \\
	v_r &=& v_r(\theta)v_\mathrm{K}, \label{eq_ss2} \\
	v_\theta &=& v_\theta(\theta)v_\mathrm{K}, \label{eq_ss3} \\
	v_\phi &=& v_\phi(\theta)v_\mathrm{K}, \label{eq_ss4} \\
	p &=& p(\theta)GMr^{-n-1}, \label{eq_ss5}
\end{eqnarray}
where $v_\mathrm{K}=\sqrt{GM/r}$ is the Keplerian velocity.
With these assumptions, Equations (\ref{eq_continuity}) and (\ref{eq_motion}) are reduced to four ODEs independent of $r$ [see equations (16)--(19) in \citealp{JW11} for the detailed forms]. However, the $r$-dependency is not immediately eliminated in Equation (\ref{eq_energy}). Here we write the equation as
\begin{equation}\label{eq_energy2}
	q_\mathrm{adv}=q_\mathrm{vis} - q_\mathrm{rad},
\end{equation}
where
$q_\mathrm{adv}$ is the advective energy transport, defined as
\begin{equation}\label{qadv}
	q_\mathrm{adv} \equiv \boldsymbol{\nabla \boldsymbol{\cdot}} (E \boldsymbol{v}) + p \boldsymbol{\nabla \boldsymbol{\cdot} v},
\end{equation}
and $q_\mathrm{rad}$ is the radiative energy transport, defined as
\begin{equation}\label{qrad}
	q_\mathrm{rad} \equiv \boldsymbol{\nabla \boldsymbol{\cdot} F}.
\end{equation}
Under self-similar assumptions, we find that $q_\mathrm{adv} \propto r^{-5/2-n}$, $q_\mathrm{vis} \propto r^{-5/2-n}$, and $q_\mathrm{rad} \propto r^{-3}$. Thus in the energy equation, we cannot eliminate the $r$-dependency as in the continuity and momentum equations, unless we set $n=1/2$. Note that the mass inflow rate obeys $\dot{M}_\mathrm{in} \propto r^{3/2-n}$ according to Equation (\ref{eq_ss1}) \& (\ref{eq_ss2}).
Theoretical studies of \cite{Begelman2012} proposed that hypercritical accretion should have $\dot{M} \propto r$, so that $n=0.5$ (but see \citealp{Jiao2022} for a caveat in their work). As for numerical simulations, most works have shown that outflow becomes negligibly small close to the BH (generally for $r<10r_\mathrm{S}$ while the exact location is dependent on the specific simulation), which is not the region we study here. Outside this region, some simulations have presented the power-law fits of the density or mass inflow rate profiles. For example, \cite{Ohsuga2005} performed 2D radiation-hydrodynamic (RHD) simulations of super-Eddington accretion flows and found that the mass accretion rate is roughly in proportion to radius in the region which achieves a quasi-steady state, so that $n \sim 0.5$. \cite{Yang2014} expanded their work with an improved treatment of the viscous stress and found $n \approx 0.55$ (denoted $p$ in their paper). \cite{Kitaki2018} performed a series of 2D RHD simulations with various BH masses and accretion rates for super-Eddington accretion covering the space of 2--3000$r_\mathrm{S}$, and achieved a relatively large $r_\mathrm{qss} \sim 200r_\mathrm{S}$ inside which the simulation establishes a quasi-steady state, and they found $\rho \propto r^{-0.73}$ so that $n=0.73$.
On the other hand, the three-dimensional general relativistic radiation-magnetohydrodynamic (3D GRRMHD) simulation performed by \cite{McKinney2014} found that $\dot{M} \propto r^{1.7}$ so that $n=-0.2$, though their simulation is for super-Eddington accretion flow onto a fast-spinning BH ($a/M = 0.9375$) and the power-law fit includes the region that has not achieved inflow equilibrium, so their result may not be applicable to our study here.
Overall, we can see that the value of $n$ is still an open question. Note that while simulations with more physically-realistic assumptions keep being developed, they are also more expensive in computation and generally achieve a quasi-steady state with inflow equilibrium at a smaller radius (e.g., the simulation in \citealp{McKinney2014} has $r_\mathrm{qss} \sim 14 r_\mathrm{S}$), so their power-law fits of radial density profiles are not necessarily better.
In this paper, we will just set $n=1/2$, and leave the discussion of $n$ values to future works.

For $n=1/2$, the full set of equations becomes a set of ODEs independent of $r$, which can then be solved with proper boundary conditions and three input parameters [$\alpha$, $M$, $\rho(\pi/2)$]. Here $\alpha$ is the viscous parameter, $M$ is the mass of the central BH, and $\rho(\pi/2)$ is the value of $\rho(\theta)$ in cgs units on the equatorial plane in Equation (\ref{eq_ss1}), which is determined eventually by the accretion rate. So the input parameters are essentially identical to conventional height-integrated disk models, which are ($\alpha$, $M$, $\dot{M}$). Compared with the input parameters in our previous self-similar solution \citep{JW11}, which are ($\alpha$, $f$, $\gamma_{\mathrm{equ}}$, $n$), the advective factor, $f = q_\mathrm{adv}/q_\mathrm{vis}$, is no longer a constant input parameter, but instead can be calculated from the equations; the equivalent heat capacity ratio, $\gamma_{\mathrm{equ}}$, is omitted, as gas pressure is negligible in super-Eddington accretion flows; the radial density index, $n$, is now set to be 1/2. On the other hand, the calculation of radiative flux involves the actual values of density and radius, unlike \cite{JW11} where only the relative value of density matters, so $M$ and $\rho(\pi/2)$ are now added as two new input parameters.

\subsection{Boundary Conditions}\label{bcondition}
We assume that the accretion flow has reflection symmetry about the equatorial plane, so that
\begin{equation}\label{eq_boundary}
	\theta=\frac{\pi}{2}:
	v_\theta=\frac{\partial\rho}{\partial\theta}=\frac{\partial p}{\partial\theta}=\frac{\partial v_r}{\partial\theta}=\frac{\partial v_\phi}{\partial\theta}=0.
\end{equation}
As mentioned in Section \ref{eqs}, we also set $\rho(\pi/2)$ as a boundary condition (given as an input parameter), which is eventually determined by the mass accretion rate of the flow. Note that in the energy Equation (\ref{eq_energy}), the term $\nabla \boldsymbol{\cdot} \boldsymbol{F}$ contains 2nd-order derivatives of $p$, which must be dealt with. Under self-similar assumptions, the radiative flux $\boldsymbol{F}$ is expressed as
\begin{equation}\label{eq_F}
	\boldsymbol{F} = \frac{(1+n)GMc p(\theta)}{r^2 \kappa \rho(\theta)} \hat{r} - \frac{GMc}{r^2 \kappa \rho(\theta)} \frac{\partial p(\theta)}{\partial \theta} \hat{\theta},
\end{equation}
neglecting gas pressure. We can see that $r^2 F_r$ actually becomes independent of $r$, so when calculating $\boldsymbol{\nabla \boldsymbol{\cdot} F}$, the contribution of $\boldsymbol{F}_r$ is 0. Physically this represents that the radiative flux flowing into and out of a fixed region in $r$ direction just cancels out each other. Thus we only need to consider $p''(\theta)$ as the 2nd-order term in the energy equation. Here we define $z_p(\theta) \equiv p'(\theta)$, so that $p''(\theta) = z_p'(\theta)$, transferring the 2nd-order differential equation into two 1st-order ones, so that the problem becomes a set of six ODEs with six unknowns.

To start the ODE solver from the equatorial boundary, we need to calculate the boundary values of the six unknowns, namely $\rho(\pi/2)$, $v_r(\pi/2)$, $v_\theta(\pi/2)$, $v_\phi(\pi/2)$, $p(\pi/2)$, and $z_p(\pi/2)$. The value of $\rho(\pi/2)$ is given as an input parameter, and Equation (\ref{eq_boundary}) provides the other five boundary conditions, although some of them are derivatives of the unknowns on the equatorial boundary. Ideally, we should be able to calculate the equatorial values of the six unknowns with the six ODEs at $\theta=\pi/2$.
However, the momentum equation in $\theta$ direction, which is 
\begin{equation}\label{eq_momentum2}
	\rho(\theta)\left[v_\theta(\theta)\left(v_r(\theta)+2v_\theta'(\theta)\right)-2\cot{\theta}v_\phi(\theta)^2\right]+2p'(\theta)=0
\end{equation}
under self-similar assumptions, becomes the form "0=0" on the equatorial plane. To solve this problem, we differentiate this equation with respect to $\theta$, which yields
\begin{eqnarray}\label{eq_dmomentum2}
	\rho(\theta) [v_\theta(\theta) \left(v_r'(\theta)+2 v_\theta''(\theta)\right)+v_\theta'(\theta) \left(v_r(\theta) + 2 v_\theta'(\theta)\right) \nonumber \\
	-4 \cot (\theta) v_\phi(\theta) v_\phi'(\theta)+2 \csc^2(\theta) v_\phi(\theta)^2] \nonumber \\
	+\rho'(\theta) \left[v_\theta(\theta ) \left(v_r(\theta) + 2 v_\theta'(\theta)\right)-2 \cot (\theta) v_\phi(\theta)^2\right]+2 z_p'(\theta)=0.
\end{eqnarray}
As we have $v_\theta''(\pi/2)=0$ according to the reflection symmetry, Equation (\ref{eq_dmomentum2}) can be used in the place of Equation (\ref{eq_momentum2}) to solve for values of the six unknowns on the equatorial boundary. Note that Equation (\ref{eq_dmomentum2}) is only used for calculating the boundary values of the unknowns, and we still use Equation (\ref{eq_momentum2}) for the ODE solver, so the problem is still a set of six first-order ODEs and the equatorial boundary conditions are sufficient for solving the problem.

\section{NUMERICAL RESULTS}\label{result}

\subsection{The Inflow-Outflow Solution}\label{f_mean}
\begin{figure}
	\includegraphics[width=\columnwidth]{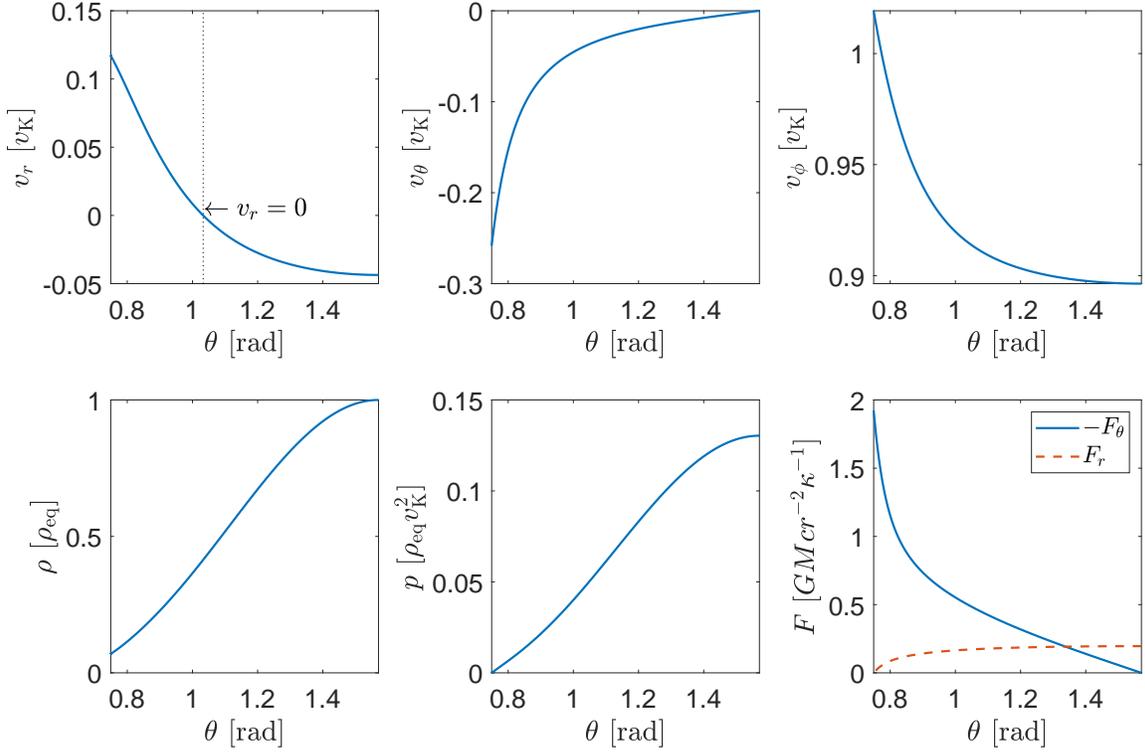}
	\caption{Self-similar solution for $\alpha=0.1$, $M=10M_\odot$, and $\rho(\pi/2)=0.5$ g cm$^{-2.5}$ [for $n=1/2$; $\rho(\pi/2)$ is a coefficient defined in our calculation, see Equation (\ref{eq_ss1})], which corresponds to the density on the equatorial plane $\rho_\mathrm{eq}=9.2\times 10^{-6}$ g cm$^{-3}$, the mass inflow rate $\dot{M}_\mathrm{in}=6800 L_\mathrm{Edd}/c^2$, and the mass outflow rate $\dot{M}_\mathrm{out}=853 L_\mathrm{Edd}/c^2$ at $r=1000 r_\mathrm{S}$. Note that the normalization factor for velocities, $v_\mathrm{K}$, is defined as $\sqrt{GM/r}$, independent of $\theta$. All these normalized profiles are independent of $r$.}
	\label{fig_sol1}
\end{figure}

A typical solution of our model, with input parameters of $\alpha=0.1$, $M=10M_\odot$, and $\rho(\pi/2)=0.5$ g cm$^{-2.5}$ [for $n=1/2$, see Equation (\ref{eq_ss1})], is presented in Figure~\ref{fig_sol1}. The corresponding density on the equatorial plane (denoted $\rho_\mathrm{eq}$), mass inflow rate (denoted $\dot{M}_\mathrm{in}$), and mass outflow rate (denoted $\dot{M}_\mathrm{out}$) are $\rho_\mathrm{eq}=9.2\times 10^{-6}$ g cm$^{-3}$, $\dot{M}_\mathrm{in}=6800 L_\mathrm{Edd}/c^2$, and $\dot{M}_\mathrm{out}=853 L_\mathrm{Edd}/c^2$ respectively at $r=1000 r_\mathrm{S}$, which scale with radius in the form $\rho \propto r^{-1/2}$ and $\dot{M} \propto r$ (in a more general case, $\rho \propto r^{-n}$ and $\dot{M} \propto r^{3/2-n}$; see the discussion on mass conservation of steady solutions in Section \ref{comparison}). Here we set $1000 r_\mathrm{S}$ as the outer boundary of the self-similar region, beyond which the mass inflow and outflow rates no longer change with radius, so $6800 L_\mathrm{Edd}/c^2$ can also be regarded as the mass supply rate from the environment. The top three panels show the $\theta$ profiles of the three components of velocity, normalized by $v_\mathrm{K}=\sqrt{GM/r}$. The bottom three panels show the $\theta$ profiles of density, pressure, and radiative fluxes, from left to right, normalized by $\rho_\mathrm{eq}$, $\rho_\mathrm{eq} v_\mathrm{K}^2$, and $GMcr^{-2}\kappa^{-1}$, respectively. In the bottom right panel, the solid line represents $-F_\theta$ while the dashed line represents $F_r$. The values of $F_\theta$ are negative as they are in the $\theta$-decreasing direction towards the polar axis. Note that all these profiles are independent of $r$.

\begin{figure}
	\includegraphics[width=\columnwidth]{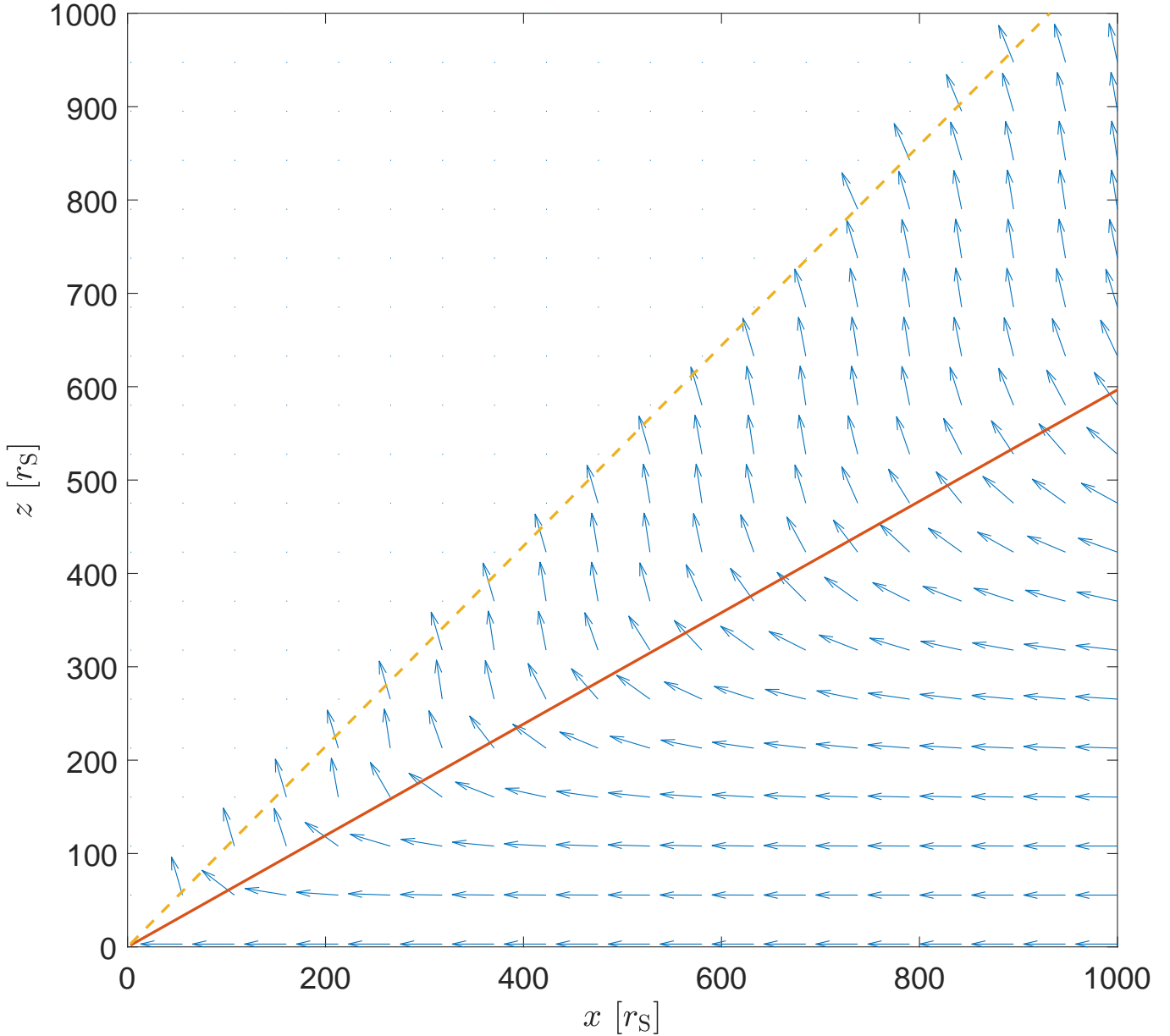}
	\caption{Velocity field diagram for the solution presented in Figure~\ref{fig_sol1}. The abscissa is $x=r\sin\theta$ and the ordinate is $z=r\cos\theta$. The arrows denote the directions of the velocity vectors in the $r\theta$ plane. The solid line represents where the radial velocity $v_r=0$ ($\theta=1.03$ rad), and the dashed line represents the calculation boundary of our self-similar solution ($\theta=0.75$ rad).}
	\label{fig_vfd}
\end{figure}

We can see that the accretion flow consists of an inflow region around the equatorial plane and an outflow region beyond the inflow, with $v_r=0$ at a certain inclination angle (for the solution presented in Figure~\ref{fig_sol1}, $\theta=1.03$ rad) between the inflow and the outflow. 
This can be more clearly seen in the velocity field diagram for the solution, Figure \ref{fig_vfd}, where the inflow region is below the solid line (due to reflection symmetry, it is actally within 0.54 rad around the equatorial plane) and the outflow region is between the solid and the dashed lines. The dashed line represents the calculation boundary of our self-similar solution, which is further explained as follows.

The calculation starts from the equatorial plane, moves towards the polar axis, and stops at the inclination where density and pressure drops very low and numerical error begins to rise so steeply that even reducing the integration step to machine accuracy fails to find a next-step solution within the error control setting. We call this inclination angle the calculation boundary of our solution. Mathematically, the calculation boundary represents the place where the ODE solver cannot proceed due to a steep increase in numerical errors, and should be distinguished from the equatorial bounday which provides boundary conditions for our calculation. We have tried different ODE solvers (see, e.g., \citealt{Hosea1996, Shampine1997, Shampine1999, Shampine2002}) with different error control settings, and find that the calculation boundaries and the solutions are basically the same. For example, when we use the same ODE solver, changing the relative error tolerance from $10^{-12}$ to $10^{-3}$ only changes the calculation boundary for less than $10^{-6}$ rad.
Thus we expect that this calculation boundary is intrinsic rather than dependent on the numerical method. 
The physical interpretation is that self-similarity cannot well describe the whole space (which it should not due to mass conservation, see Section \ref{comparison} for more details), so that above the self-similar accretion flow that our solution describes, there exists non-self-similar flow in the region around the polar axis. When our self-similar solution approaches the boundary between the self-similar and the non-self-similar regions, numerical errors will inevitably undergo a steep rise, regardless of the ODE solver and the error control setting that we choose. So the calculation boundary of our solution is a good indication of the physical boundary between the self-similar and the non-self-similar regions in the accretion flow, which is also the surface of the self-similar accretion flow that we solves.  Nevertheless, our solution solves the structure of the accretion flow in the self-similar region with a good numerical accuracy, and our analyses below based on this solution should be trustworthy within the framework of our assumptions.
We have also calculated the optical depth $\tau$ from the surface, and find that the position of $\tau=1$ is very close to the surface ($\Delta\theta=0.01$ rad at $r=10r_\mathrm{S}$, and even smaller for larger radius). So the assumption of optical thickness is well held in most of the solution, even when the optical depth in the non-self-similar region is not counted.

\begin{figure}
	\includegraphics[width=\columnwidth]{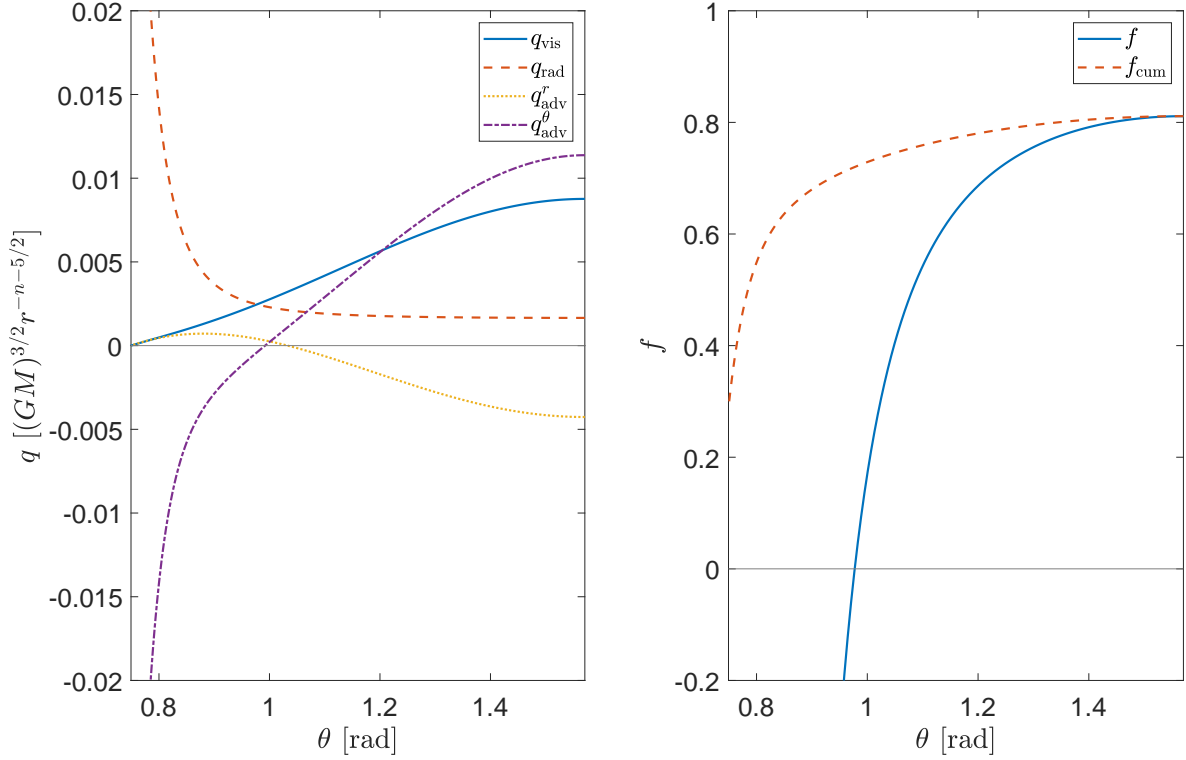}
	\caption{
		Heating/cooling rates normalized by $(GM)^{3/2} r^{-n-5/2}$ in the left panel and the advective factor in the right panel, for the solution presented in Figure~\ref{fig_sol1}. In the left panel, the solid line represents the heating caused by viscous dissipation, $q_\mathrm{vis}$, which is always positive; the dashed line represents the radiative energy transport, $q_\mathrm{rad}$; the dotted line represents the advective energy transport in $r$ direction, $q_\mathrm{adv}^r$; the dot-dashed line represents the advective energy transport in $\theta$ direction, $q_\mathrm{adv}^\theta$. The values of $q_\mathrm{rad}$, $q_\mathrm{adv}^r$, and $q_\mathrm{adv}^\theta$, can be either positive or negative, while positive values represent energy loss (cooling) and negative values represent energy gain (heating), and the sum of these values should equal $q_\mathrm{vis}$ in order to maintain energy conservation. In the right panel, the solid line represents the local advective factor, $f$, while the dashed line represents the cumulative advective factor, $f_\mathrm{cum}$. The grey solid line in each panel indicates where the function values equal 0.}
	\label{fig_qf}
\end{figure}

It is also worth noticing in Figure~\ref{fig_sol1} that the rotational speed $v_\phi$ is only $0.8964 v_\mathrm{K}$ on the equatorial plane and becomes super-Keplerian at higher latitudes. The reason of $v_\phi < v_\mathrm{K}$ near the equatorial plane is that there is significant pressure gradient pointing outward in the $r$ direction, so that the required centrifugal force to resist the gravity is reduced. We also calculated the slim disk model solution for the mass supply rate $6800 L_\mathrm{Edd}/c^2$, and found that the value of $v_\phi/v_\mathrm{K}$ on the equatorial plane inside $1000 r_\mathrm{S}$ is in the range 0.6480--0.8995, so this is to be expected in conventional models for super-Eddington accretion. The equatorial value of $v_\phi$ in our solution also agrees with numerical simulations which have discussed this value \citep[e.g.,][]{Ohsuga2005,OM2011}. Note that while in figure 5 of \cite{Jiang2014} $v_\phi$ is very close to $v_\mathrm{K}$ outside of 10$r_\mathrm{S}$, their $v_\phi$ is a  vertically averaged value, and cannot be compared directly with our equatorial value of $v_\phi$. The reason that $v_\phi$ could become super-Keplerian at higher latitudes is related to the conservation of angular momentum and the geometry of spherical coordinates (i.e., the distance to the axis of rotation decreases as $\theta$ decreases), which is further discussed in Section \ref{AngularM}.

Compared with our previous self-similar solution \citep{JW11}, a large improvement is that now the model includes the calculation of the radiative energy transport $q_\mathrm{rad}$, so that the advective factor $f$ is calculated self-consistently instead of being set as an input parameter. The energy mechanisms are presented in the left panel of Figure~\ref{fig_qf}, where the solid line represents the heating rate of viscous dissipation $q_\mathrm{vis}$, the dashed line represents the radiative energy transport $q_\mathrm{rad}$, the dotted line represents the advective energy transport in $r$ direction $q_\mathrm{adv}^r$, and the dot-dashed line represents the advective energy transport in $\theta$ direction $q_\mathrm{adv}^\theta$, all of which are normalized by $(GM)^{3/2} r^{-n-5/2}$. The values of $q_\mathrm{rad}$, $q_\mathrm{adv}^r$, and $q_\mathrm{adv}^\theta$ can be either positive or negative, while positive values represent energy loss (cooling) and negative values represent energy gain (heating), and the sum of these values should equal $q_\mathrm{vis}$ in order to maintain energy conservation. We can see that $q_\mathrm{rad}$ is always a cooling mechanism in our solution, so that the term "radiative cooling" is actually accurate. It also gets stronger closer to the surface, where photons can escape more easily.
Advection, on the other hand, is somewhat complex. While the radial advection in the slim disk model is an important or even dominant cooling mechanism for super-Eddington accretion, it is now a heating mechanism in the inflow of our solution. As shown in Equation (\ref{qadv}), the advective energy transport can be divided to contributions of advected internal energy (here mostly radiative energy), $\boldsymbol{\nabla \boldsymbol{\cdot}} (E \boldsymbol{v})$, and compression heating/expansion cooling, $p \boldsymbol{\nabla \boldsymbol{\cdot} v}$. Under self-similar assumptions, their contributions in $r$ direction can be further expressed as  \citep[see][for more details]{Jiao2022}
\begin{eqnarray}
	q_\mathrm{adv}^r = q_\mathrm{adv}^{r1}+q_\mathrm{adv}^{r2} = 3 c_1 (1-n) v_r(\theta ), \label{eq_qr} \\
	q_\mathrm{adv}^{r1} = \frac{3}{2} c_1 (1-2 n) v_r(\theta ), \label{eq_qr1} \\
	q_\mathrm{adv}^{r2} = \frac{3}{2} c_1 v_r(\theta ), \label{eq_qr2} \\
\end{eqnarray}
where
\begin{equation}\label{eq_c1}
	c_1=p \Omega_\mathrm{K}=p(\theta) (GM)^{3/2} r^{-n-5/2},
\end{equation}
and $q_\mathrm{adv}^{r1}$ and $q_\mathrm{adv}^{r2}$ denote the advected internal energy and compression heating/expansion cooling in $r$ direction, respectively.
As we set $n=1/2$ in the solution, we actually have $q_\mathrm{adv}^{r1}=0$, which means that the incoming and outgoing internal energy in $r$ direction for a fixed region in the rest frame just cancel out each other and do not influence the energy balance. The radial advection is thus dominated by compression heating in the inflow and expansion cooling in the outflow, as shown by the dotted line in the left panel of Figure~\ref{fig_qf}.
The advection in $\theta$ direction $q_\mathrm{adv}^\theta$, on the other hand, acts as a cooling mechanism in the inflow and carries the left-over entropy that cannot be cooled by the local radiation upwards, which is eventually cooled by radiation or carried away by the outflow (i.e. cooled by expansion) at higher latitude. In the outflow, $q_\mathrm{adv}^r$ becomes a cooling mechanism due to expansion. The advection in $\theta$ direction $q_\mathrm{adv}^\theta$ switches to heating when the sum of radiative and expansion cooling becomes large enough to offset the viscous heating, which happens somewhere in the outflow. The radiative cooling $q_\mathrm{rad}$ gets much stronger closer to the surface, where $q_\mathrm{adv}^\theta$ becomes the main heating mechanism, which is essentially the accumulated entropy that cannot be sufficiently cooled in the inflow and the lower part of the outflow.

The $\theta$ profiles of the local advective factor $f$ (solid line) and the cumulative advective factor $f_\mathrm{cum}$ (dashed line) are presented in the right panel of Figure~\ref{fig_qf}, where $f_\mathrm{cum}$ represents the ratio of advective energy transport to viscous heating integrated from the equatorial plane to the current $\theta$ and is calculated as
\begin{equation}\label{eq_f_cum}
	f_\mathrm{cum} = \frac{\iiint\limits_V q_\mathrm{adv}dV}{\iiint\limits_V q_\mathrm{vis}dV} = \frac{\int_\theta^{\frac{\pi}{2}} q_\mathrm{adv} \sin{\theta} d\theta}{\int_\theta^{\frac{\pi}{2}} q_\mathrm{vis} \sin{\theta} d\theta}.
\end{equation}
We present this value because as density and pressure drops close to 0 near the surface, the local viscous heating $q_\mathrm{vis}$, which is the denominator in the calculation of $f$, also approaches 0 and causes $f$ to become unusually large in absolute value, so that $f_\mathrm{cum}$ actually describes the energy mechanism better. We can see that as $\theta$ decreases, radiation gets stronger, and consequently $f_\mathrm{cum}$ keeps decreasing, indicating that radiation becomes more important in the total cooling. The mean advective factor of the accretion flow (denoted $\bar{f}$), which is $f_\mathrm{cum}$ calculated at the surface, is only 0.29, much lower than the local advective factor on the equatorial plane, 0.81, and also smaller than that expected by the slim disk model, which is generally advection-dominated ($>0.5$).

We have also made multiple calculations with different input parameters, and find that outflow always exists as long as the input parameters are reasonable values under our assumptions. 
Note that our model itself does not consider the effects of gas pressure and absorption opacity, so that we have to rely on other theoretical studies or numerical simulations to ensure that neglecting these effects are feasible. We have discussed in Section \ref{eqs} that these assumptions are justified in the inner region of a super-Eddington accretion flow, ranging from the inner edge of the flow to several hundred or thousand times of Schwarzschild radius $r_\mathrm{S}$, depending on the accretion rate. So a simple way to ensure that our model is applicable is to ensure that the flow is super-Eddington. For observations, if the observed source is known to have an Eddington ratio larger than 1, then the application of our model is feasible for its inner region.
For theoretical studies with a given set of input parameters, we can calculate the effective accretion rate of our model as
\begin{equation}\label{eq_inflow}
	\dot{M}_{\mathrm{eff}} = 2\int_{\theta_\mathrm{s}}^{\frac{\pi}{2}} 2 \pi r^{2} \sin \theta \rho v_{r} d \theta = 4\pi \sqrt{GM}
	r \int_{\theta_\mathrm{s}}^{\frac{\pi}{2}}
	v_r(\theta)\rho(\theta)\sin{\theta}d\theta,
\end{equation}
where $\theta_\mathrm{s}$ is the value of $\theta$ at the surface of our solution. As the black hole mass is known as an input parameter, we can calculate $\dot{M}_\mathrm{Edd}$ (here we adopt the definition of $\dot{M}_\mathrm{Edd}=16L_\mathrm{Edd}/c^2$, see, e.g., \citealt{Ab2010}; we try to avoid a direct calculation of the radiative efficiency in our model, which would require additional assumptions, see Section \ref{F&L}) and compare these two values. The application of our model would be well justified in the inner region of the accretion flow when $\dot{M}_{\mathrm{eff}} > \dot{M}_\mathrm{Edd}$. 
Note that $\dot{M}_{\mathrm{eff}} \propto r$, and the eventual mass accretion rate that falls into the accretor can be estimated as $\dot{M}_{\mathrm{eff}}$ at the inner edge of the self-similar region, inside which outflow becomes negligible. While we do not know the exact place of the inner edge of the self-similar region, it is generally considered to be away from the inner boundary of the accretion flow \citep{NY94,Narayan97,Jiao15}, so we can be certain that this inner edge is larger than the innermost circular orbit (ISCO) at $r_\mathrm{ISCO}=3r_\mathrm{S}$. Therefore, $\dot{M}_{\mathrm{eff}}$ calculated at $r_\mathrm{ISCO}$ is an lower limit (hereafter denoted $\dot{M}_{\mathrm{ll}}$), and if it is larger than $\dot{M}_\mathrm{Edd}$, the accretion flow can surely be considered super-Eddington (the real accretion rate is likely much larger than this lower-limit value).

The above criterion also incorporates the effects of changing input parameters. 
For example, with fixed values of $\alpha=0.1$ and $M=10M_\odot$, the solution for $\rho(\pi/2)=0.46$ g cm$^{-2.5}$ has a lower-limit of the effective accretion rate $\dot{M}_{\mathrm{ll}}=\dot{M}_\mathrm{Edd}$, and the corresponding boundary between inflow and outflow (hereafter denoted $\theta_0$) is $\theta_0=1.036$ rad, the surface of the solution is $\theta_\mathrm{s}=0.764$ rad, and the mean advective factor is $\bar{f}=0.266$. As a comparison, the solution presented in Figure \ref{fig_sol1} for $\rho(\pi/2)=0.5$ g cm$^{-2.5}$ has $\dot{M}_{\mathrm{ll}} = 1.12 \dot{M}_\mathrm{Edd}$, $\theta_0=1.03$ rad, $\theta_\mathrm{s}=0.75$ rad, and $\bar{f}=0.29$.
We can see that for $\rho(\pi/2)=0.46$ g cm$^{-2.5}$, the sizes of the inflow and the outflow regions and the mean advective factor have all decreased, corresponding to a reduced accretion rate. The effective accretion rate is not linear to $\rho(\pi/2)$, because the sizes of the inflow and the outflow regions also decrease as $\rho(\pi/2)$ decreases.

The influence of changing black hole mass $M$ on the solution is not so intuitive as changing $\rho(\pi/2)$. When we compare the structure of different solutions, we usually study the region with the same values of $r$ in unit of $r_\mathrm{S}$ ($\equiv 2GM/c^2$). Note that the value of $r$ in Equation (\ref{eq_inflow}) is in cgs units, so that $r \propto r_\mathrm{S} \propto M$. Then, according to Equation (\ref{eq_inflow}), we can see that $\dot{M}_{\mathrm{eff}} \propto M^{3/2}$ when the differences in the profiles of $v_r(\theta)$ and $\rho(\theta)$ are ignored. On the other hand, the Eddington accretion rate $\dot{M}_\mathrm{Edd} \propto M$, so eventually the accretion rate in unit of $\dot{M}_\mathrm{Edd}$ is $\dot{m}=\dot{M}_{\mathrm{eff}} /\dot{M}_\mathrm{Edd} \propto M^{1/2}$ (when the change in vertical structure is considered, $\dot{m}$ should change faster than $M^{1/2}$, while a positive correlation still holds). Thus when $\alpha$ and $\rho(\pi/2)$ are fixed, decreasing the black hole mass $M$ also decreases the accretion rate (in unit of $\dot{M}_\mathrm{Edd}$). For example, with $\alpha=0.1$ and $\rho(\pi/2)=0.5$ g cm$^{-2.5}$, the solution for $M=5M_\odot$ has $\dot{M}_{\mathrm{ll}} = 0.69 \dot{M}_\mathrm{Edd}$, $\theta_0=1.05$ rad, $\theta_\mathrm{s}=0.84$ rad, and $\bar{f}=0.18$. We can see that the sizes of the inflow and the outflow regions and the mean advective factor have all decreased again, corresponding to a reduced accretion rate (in unit of $\dot{M}_\mathrm{Edd}$).

\subsection{Comparison with the Self-Similar Solution in \cite{Zeraatgari2020}}\label{comparison}
\cite{Zeraatgari2020} also established a self-similar solution for super-Eddington accretion. Compared with our solution, they considered all components of the viscous stress tensor, so that more boundary conditions are required and they applied the symmetric boundary conditions on both the equatorial plane and the polar axis. Consequently, the self-similar assumptions are adopted in the full space in their solution.
Note that for a steady solution, the total accretion rate should be constant, which with reflection symmetry can be expressed as
\begin{equation}\label{eq_mdot}
	\dot{M}_{\mathrm{total}} \equiv 2\int_0^{\frac{\pi}{2}} 2 \pi r^{2} \sin \theta \rho v_{r} d \theta = \mathrm{const}.
\end{equation}
If self-similarity holds for the whole range of $\theta$, we can further get,
\begin{equation}\label{eq_mdot_ss}
	\dot{M}_\mathrm{total} = 4\pi \sqrt{GM}
	r^{\frac{3}{2}-n} \int_0^{\frac{\pi}{2}}
	v_r(\theta)\rho(\theta)\sin{\theta}d\theta = \mathrm{const}.
\end{equation}
As \cite{Zeraatgari2020} also set $n=1/2$, their solution must have $\dot{M}_\mathrm{total}=0$. However, the direct applications of super-Eddington accretion models in astronomy are to observations with super-Eddington luminosities, such as ULXs, whose energy eventually comes from the gravitational energy of accreted matter. The solution of \cite{Zeraatgari2020} requires that the mass outflow exactly cancels the mass inflow, so that no (or negligible) mass are actually accreted, which cannot explain the high luminosities in these observations.

On the other hand, our self-similar solution does not cover the whole range of $\theta$, and we expect that there exists a non-self-similar region around the polar axis, so that the total accretion rate in our solution is
\begin{equation}\label{eq_mdot_our}
	\dot{M}_{\mathrm{total}} = \dot{M}_{\mathrm{ss}}+\dot{M}_{\mathrm{polar}},
\end{equation}
where $\dot{M}_{\mathrm{ss}}$ is the net accretion rate in the self-similar region and $\dot{M}_{\mathrm{polar}}$ is the accretion rate of the non-self-similar region around the polar axis. As $\dot{M}_{\mathrm{polar}}$ does not follow the self-similar form, our solution no longer requires a total accretion rate of 0 as in \cite{Zeraatgari2020}, so that a large total accretion rate can be allowed in our solution. The trade-off is that our solution can only describe a limited range of $\theta$. 
The direction of velocity on the surface of our solution in Figure~\ref{fig_sol1} is still pointing upward and outward, so that there is outflow blowing from the surface into the non-self-similar region, and we speculate that the accretion flow in the non-self-similar region is also in the form of outflow, which helps to maintain a constant total accretion rate. This configuration is consistent with the structure obtained in numerical simulations of super-Eddington accretion (see references in Section \ref{intro}), though the exact structure in the non-self-similar region is beyond our calculation.

Another difference is that we only consider the $r\phi$-component of the viscous stress tensor in the calculation, while \cite{Zeraatgari2020} considered all the components. However, the inclusion of other components of the viscous stress does not change the formula of advective energy transport, and only changes the formula of the viscous heating, which would still be a heating term anyway, so the general results in this paper should still hold.

\subsection{Conservation of Angular Momentum}\label{AngularM}
In the classical picture of viscous accretion disk theory, the viscosity caused by differential rotation transfers angular momentum outward and disk material inward \citep{LP1974}. When we focus on a fixed region in the rest frame, the viscous torque reduces the angular momentum in the region, but the material flowing into the region carries more angular momentum than that flowing out of the region, so that the angular momentum in the fixed region can be conserved for a steady solution. We can regard this gain (or loss) of angular momentum due to flow motion as the advection of angular momentum. 
In conventional height-integrated models, vertical motion is ignored, so that the radial advection of angular momentum exactly balances the viscous torque, and the angular momentum per unit mass relative to the axis of rotation would remain constant in the vertical direction. For a solution in cylindrical coordinates, this would result in a constant $v_\phi$ in $z$ direction. However, in spherical coordinates, the angular momentum per unit mass is $l=r \sin\theta v_\phi$, and a constant $l$ means that $v_\phi$ would increase as $\theta$ decreases in the form of $v_\phi \propto \csc\theta$, becasue the distance to the axis of rotation is proportional to $\sin\theta$. The real case is more complicated when vertical motion and advection are considered, especially in the outflow region where the radial advection also tends to decrease the angular momentum in the fixed region, and the conservation of angular momentum is upheld by the vertical advection.

\begin{figure}
	\includegraphics[width=\columnwidth]{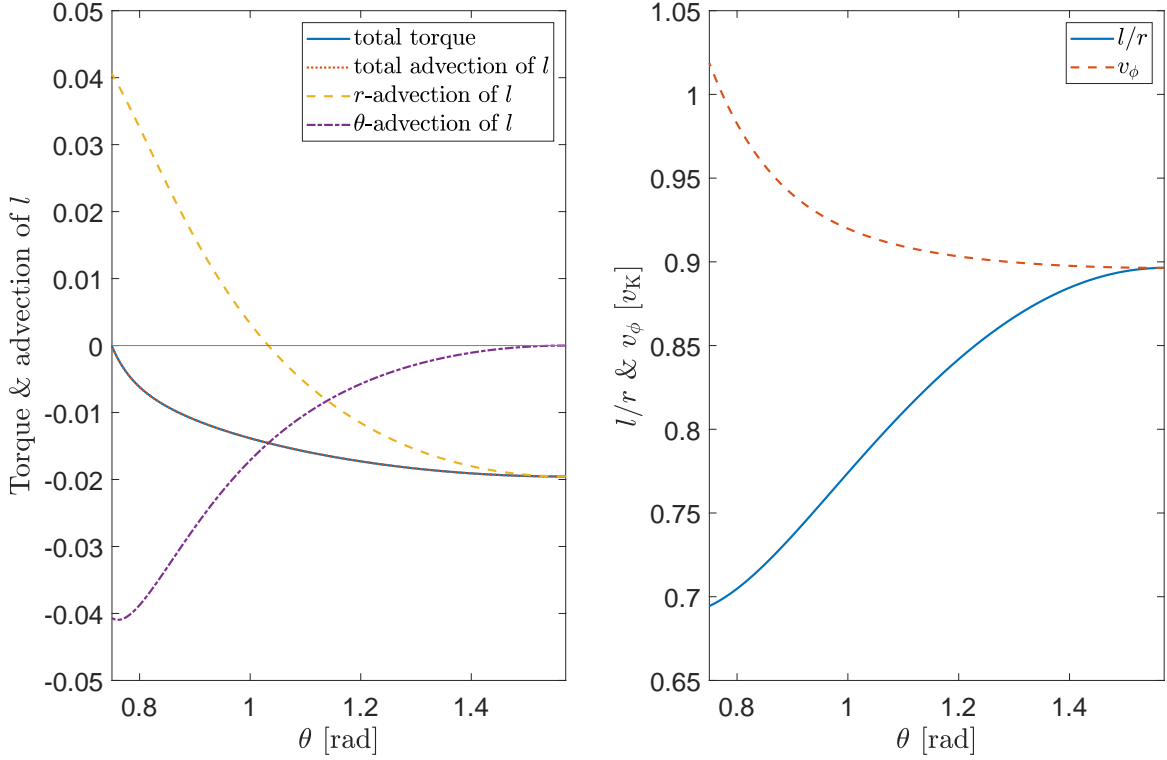}
	\caption{
		Left panel: total torque and advection of angular momentum, normalized by $GM/r$; right panel: the comparison of angular momentum per unit mass normalized by $r$, $l/r=\sin\theta v_\phi$, and the rotational velocity $v_\phi$, for the solution presented in Figure~\ref{fig_sol1}. In the left panel, the solid line represents the total torque, the dotted line represents the total advection of angular momentum, and the dashed and dot-dashed lines represent the advection of angular momentum in $r$ and $\theta$ directions, respectively. Positive values of advection represent loss, and negative values represent gain, of angular momentum. The total torque is always negative, indicating that viscosity tends to reduce the angular momentum for a fixed region. The solid and dotted lines coincide with each other, indicating that angular momentum conservation is maintained in our steady solution. The grey solid line indicates where the function values equal 0. In the right panel, the solid line represents $l/r$ while the dashed line represents $v_\phi$.}
	\label{fig_angM}
\end{figure}

To study the conservation of angular momentum relative to the polar axis (which is the axis of rotation here), we calculate $\boldsymbol{r} \times$ Equation (\ref{eq_motion}), project the result onto the polar axis, and divide the equation by $\rho$ (to study the angular momentum per unit mass), 
which eventually yields 
\begin{equation}\label{eq_angM}
	\sin{\theta} v_r \left(v_\phi + r\frac{\partial v_\phi}{\partial r} \right) + \sin\theta v_\theta \left(\cot\theta v_\phi + \frac{\partial v_\phi}{\partial \theta} \right) = \frac{1}{\rho} r\sin{\theta}N_\phi,
\end{equation}
where $N_\phi$ is the $\phi$-component of the viscous force
\begin{equation}\label{eq_Nphi}
	N_\phi=\frac{1}{r^3} \frac{\partial}{\partial r}\left(r^3 t_{r \phi}\right).
\end{equation}
The RHS of Equation (\ref{eq_angM}) is the total torque per unit mass, contributed completely by the viscous force as gravity and pressure gradient exert no torque about the polar axis due to axisymmetry, while the two terms on the LHS are the radial and vertical advection of angular momentum, respectively. The two components of the advection of angular momentum and the viscous torque per unit mass, calculated from our solution, are displayed in the left panel of Figure~\ref{fig_angM}. The solid line represents the total torque, the dotted line represents the total advection of angular momentum, and the dashed and dot-dashed lines represent the advection of angular momentum in $r$ and $\theta$ directions, respectively. The advection of angular momentum is calculated as the difference between the outgoing and the incoming angular momentum, so that positive values here represent loss, and negative values represent gain, of angular momentum. The total torque is always negative, indicating that viscosity tends to reduce the angular momentum in the fixed region. The total advection of angular momentum coincides with the total torque, indicating that the angular momentum conservation is maintained, as expected for a steady solution. The radial advection starts out as angular momentum gain on the equatorial plane, decreases as $\theta$ decreases, and becomes angular momentum loss in the outflow region, where the angular momentum conservation is upheld solely by the vertical advection. The vertical advection always increases the angular momentum in the fixed region, which indicates that the angular momentum per unit mass tends to decrease in the direction of $v_\theta$. This can be clearly seen in the right panel of Figure \ref{fig_angM}, where the solid line represents angular momentum per unit mass normalized by $r$, $l/r=\sin\theta v_\phi$, and the dotted line represents $v_\phi$, shown here as a comparison. We can see that although the angular momentum per unit mass decreases as $\theta$ decreases for a fixed radius, $v_\phi$ keeps increasing as the distance to the axis of rotation decreases faster than the angular momentum.

\subsection{Outflow-Driving Mechanism}\label{outflowDM}
We expand the equation of motion, Equation (\ref{eq_motion}), in the $r$ and $\theta$ directions, and get
\begin{equation}\label{eq_motionr}
	v_r \frac{\partial v_r}{\partial
		r}+\frac{v_{\theta}}{r}(\frac{\partial v_r}{\partial
		\theta}-v_\theta)-\frac{v_\phi^2}{r}=-\frac{GM}{r^2}-\frac{1}{\rho} \frac{\partial
		p}{\partial r},
\end{equation}
\begin{equation}\label{eq_motiont}
	v_r \frac{\partial v_\theta}{\partial
		r}+\frac{v_{\theta}}{r}(\frac{\partial v_\theta}{\partial
		\theta}+v_r)-\frac{v_\phi^2}{r} \cot \theta=-\frac{1}{\rho r} \frac{\partial
		p}{\partial \theta},
\end{equation}
where the last terms on the LHS are the $r$ and $\theta$ components of the centrifugal force per unit mass, $f_\mathrm{cen}=v_\phi^2/(r\sin\theta)$, whose direction is perpendicular to the axis of rotation. Note that in spherical coordinates, gravity has no component in the vertical ($\theta$) direction while centrifugal force has a component, $\cos\theta \cdot f_\mathrm{cen}=v_\phi^2\cot\theta/r$, and the motion in the vertical ($\theta$) direction is thus determined by the pressure gradient and the centrifugal force. This is quite different from the analyses of accretion flow solutions in cylindrical coordinates, where the motion in the vertical ($z$) direction is determined by the $z$-components of gravity and pressure gradient while centrifugal force has no influence.
As we ignored gas pressure due to the domination of radiation pressure in our solution, the pressure gradient is actually the radiative force, whose components in $r$ and $\theta$ directions agree with the profiles of $F_r$ and $F_\theta$ (the bottom-right panel of Figure \ref{fig_sol1}).

\begin{figure}
	\includegraphics[width=\columnwidth]{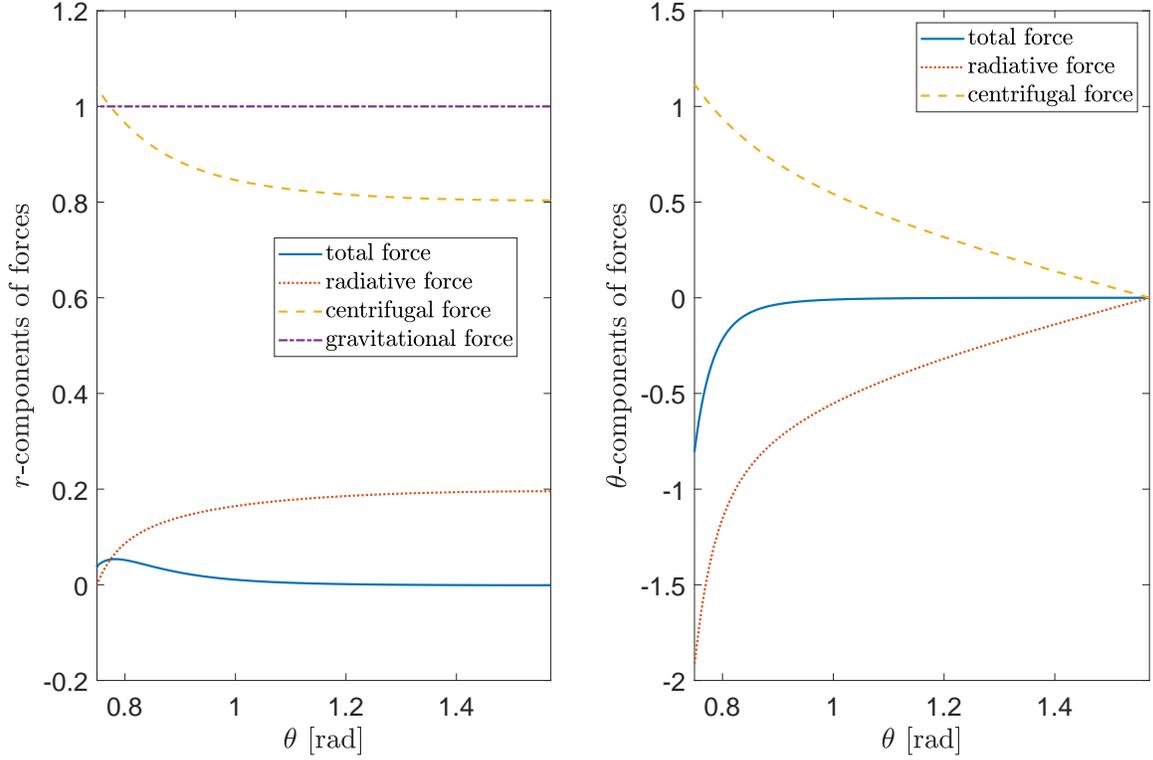}
	\caption{
		Components of forces in $r$ direction (left panle) and $\theta$ direction (right panel), for the solution presented in Figure~\ref{fig_sol1}. In each panel, the solid line represents the total force, the dotted line represents the radiative force (i.e., pressure gradient), the dashed line represents the centrifugal force, and the dot-dashed line represents the gravitational force. The directions are in typical definition of spherical coordinates (except for the gravitational force which we present the absolute value here), so that positive values in $r$ direction points outward away from the accretor, and positive values in $\theta$ direction points downward away from the polar axis.
		Note that in spherical coordinates, the gravitational force has no component in $\theta$ direction while the centrifugal force has components in both $r$ and $\theta$ directions.
	}
	\label{fig_force}
\end{figure}

Components of forces (per unit mass) in $r$ and $\theta$ directions are shown in Figure~\ref{fig_force}. In each panel, the solid line represents the total force, the dotted line represents the radiative force (i.e., pressure gradient), the dashed line represents the centrifugal force, and the dot-dashed line represents the gravitational force. The directions are in typical definition of spherical coordinates (except for the gravitational force which we present the absolute value here), so that positive values in $r$ direction points outward away from the accretor, and positive values in $\theta$ direction points downward away from the polar axis.

In $r$ direction, the motion is determined by the combined effects of radiative, centrifugal, and gravitational forces.
The total force in $r$ direction is marginally negative (i.e., pointing inward) near the equatorial plane, but it needs to provide the inward acceleration of $v_r$ in $r$ direction, which means that the first term in Equation (\ref{eq_motionr}) must be negative. The total effect for the acceleration of $v_r$ in $\theta$ direction is thus always equivalent to an outward-pointing force, i.e., $\partial v_r/\partial \theta$ is always negative (except for the equatorial plane where $\partial v_r/\partial \theta=0$ due to reflection symmetry). Therefore, as $\theta$ decreases, $v_r$ tends to increase outward, and the inward radial motion on the equatorial plane will eventually turn outward and outflow is developed, as the combination of radiative and centrifugal forces overcomes the gravity.
We can also see that the $r$-component of the radiative force maintains a relatively high level in the majority of the flow except for the region very close to the surface, and plays an important role in resisting the gravity and driving the outflow, although the dominant outward-pointing force is still the centrifugal force. This agrees with numerical simulations of super-Eddington accretion that have presented the analyses of forces in $r$ direction (e.g., fig. 5 of \citealt{OM2007}; figure 5 of \citealt{Yang2014}). Some simulations have stated that close to the polar axis, the radiative force becomes dominant in $r$ direction \citep[e.g.,][]{Ohsuga2005,Ohsuga2009,OM2007,OM2011,Yang2014,Sadowski2014}, but our solution can not reach there as the region around the polar axis is non-self-similar.

In $\theta$ direction, the radiative force always points away from the equatorial plane, as the radiation tries to escape the accretion flow and the $\theta$-component of radiative flux, $F_\theta$, always points away from the equatorial plane and keeps increasing as $\theta$ decreases. The $\theta$-component of centrifugal force, on the other hand, pushes the material in the opposite direction. The total force is always negative in $\theta$ direction (except for on the equatorial plane where the $\theta$-components of both the total force and the velocity are 0), so that $v_\theta$ starts as 0 on the equatorial plane and becomes negative (pointing away from the equatorial plane) while its absolute value keeps increasing as $\theta$ decreases, as shown in Figure~\ref{fig_sol1}. The force analysis in $\theta$ direction agrees with the numerical simulation of \cite{Yang2014} which has presented this analysis in their figure 6.

Note that the definition of outflow and consequently the discussion of outflow-driving mechanism may differ in different papers.
For example, in \cite{Jiang2014} whose calculation is performed in cylindrical coordinates, the material that escapes vertically is also considered outflow, and the outflow-driving mechanism is discussed based on vertical ($z$) components of forces (their figure 8). If we make a similar discussion in $\theta$ direction, we can also say that our flow is radiation-driven because the radiative force in $\theta$ direction is the dominant force that drives the flow away from the equatorial plane.

\subsection{Emergent Radiative Flux, Luminosity, and Radiative Efficiency}\label{F&L}
The relatively small value of mean advective factor indicates that compared with the conventional slim disk model, a larger portion of viscous heating is converted into radiation. 
As mentioned in Section \ref{f_mean}, the advection in $\theta$ direction carries the left-over entropy at lower latitudes that cannot be radiated away upward, part of which is converted into radiation at higher latitudes. This effect is not considered in the slim disk model which neglects vertical motion. However, our solution cannot resolve the structure in the non-self-similar region, and the net accretion rate in the self-similar region of our solution decreases as radius decreases, which reduces the total viscous heating. 
So the consideration of vertical motion and outflow in our model actually influences the radiative efficiency $\eta$ in two opposite ways, when it is calculated according to the mass supply rate. Thus, it is necessary to compare the emergent radiative fluxes and luminosities of our model and the slim disk model.

The radiative flux in our model has two components, $F_r$ and $F_\theta$, under axisymmetry. As shown in the bottom right panel of Figure~\ref{fig_sol1}, $F_\theta$ starts out as 0 on the equatorial plane and increases towards the polar axis as the local heating is converted into radiation, reaching its maximum value at the surface of the accretion flow (negative values represent the direction). On the other hand, $F_r$ does not contribute to the "radiative cooling" as mentioned in Section \ref{bcondition}, and can be perceived as being generated in the inner part of the accretion flow close to the central BH, which then propagates outward in $r$ direction following the inverse-square law. 
We can actually calculate the contribution of $F_r$ to the total luminosity as
\begin{equation}\label{eq_Lr}
	L_r=2\int_{\theta_\mathrm{s}}^{\frac{\pi}{2}} F_r\boldsymbol{\cdot} 2\pi r^2 \sin{\theta} \mathrm{d} \theta,
\end{equation}
where $\theta_\mathrm{s}$ is the value of $\theta$ at the surface. For the solution presented in Figure~\ref{fig_sol1}, we get $L_r=0.124L_\mathrm{Edd}$. In height-integrated models in cylindrical coordinates, such as the SSD and the slim disk, people usually calculate the radiative cooling per unit area on the equatorial plane (denoted $Q_\mathrm{rad,cyl}$ with a subscript "cyl" to distinguish from our model in spherical coordinates) as
\begin{equation}\label{eq_Qrad1}
	Q_\mathrm{rad,cyl}=\int_{-H}^{H}q_\mathrm{rad,cyl} \mathrm{d}z
\end{equation}
where $H$ is the half thickness of the disk in cylindrical coordinates, and $q_\mathrm{rad,cyl}$ represents the radiative cooling per unit volume in these models. In this way, the emergent radiative flux $F = Q_\mathrm{rad,cyl}/2$. The total luminosity is calculated as
\begin{equation}\label{eq_L1}
	L_\mathrm{cyl}=\int_{r_\mathrm{in}}^{r_\mathrm{out}} Q_\mathrm{rad,cyl}\boldsymbol{\cdot} 2\pi r \mathrm{d}r = \int_{r_\mathrm{in}}^{r_\mathrm{out}} 2F \boldsymbol{\cdot} 2\pi r \mathrm{d}r.
\end{equation}
Apparently the emergent radiative flux as well as the luminosity in height-integrated models does not consider the contribution of the radiative flux in $r$ direction, so it corresponds to the emergent $F_\theta$ at the surface (denoted $F_{\theta \mathrm{s}}$) in our model. We can also calculate the radiative cooling per unit area of our model in a similar way,
\begin{equation}\label{eq_Qrad}
	Q_\mathrm{rad}=2\int_{\theta_\mathrm{s}}^{\frac{\pi}{2}} q_\mathrm{rad} r \sin{\theta} \mathrm{d} \theta.
\end{equation}
However, the relation between $F_{\theta \mathrm{s}}$ and $Q_\mathrm{rad}$ is actually $F_{\theta \mathrm{s}} = Q_\mathrm{rad}/(2\sin{\theta_\mathrm{s}})$. The reason is that the emitting surface is smaller by a factor of $\sin{\theta_\mathrm{s}}$ compared with the height-integrated models in cylindrical coordinates. The luminosity emergent in $\theta$ direction is thus
\begin{equation}\label{eq_Ltheta}
	L_\theta=\int_{r_\mathrm{in}}^{r_\mathrm{out}} Q_\mathrm{rad}\boldsymbol{\cdot} 2\pi r \mathrm{d}r = \int_{r_\mathrm{in}}^{r_\mathrm{out}} 2F_{\theta \mathrm{s}}\sin{\theta_\mathrm{s}} \boldsymbol{\cdot} 2\pi r \mathrm{d}r,
\end{equation}
which is basically the same formula for $Q_\mathrm{rad}$ as Equation (\ref{eq_L1}). Note that self-similar assumptions are only applicable to a certain range of radii in the accretion flow, and does not hold in, e.g., the region very close to the BH where general relativistic effects are strong, while the contribution to luminosity there cannot be neglected. So the luminosity emergent in $\theta$ direction cannot be obtained in this paper without further assumptions. However, it can still be compared indirectly by comparing the radiative fluxes in the self-similar region.

\begin{figure}
	\includegraphics[width=\columnwidth]{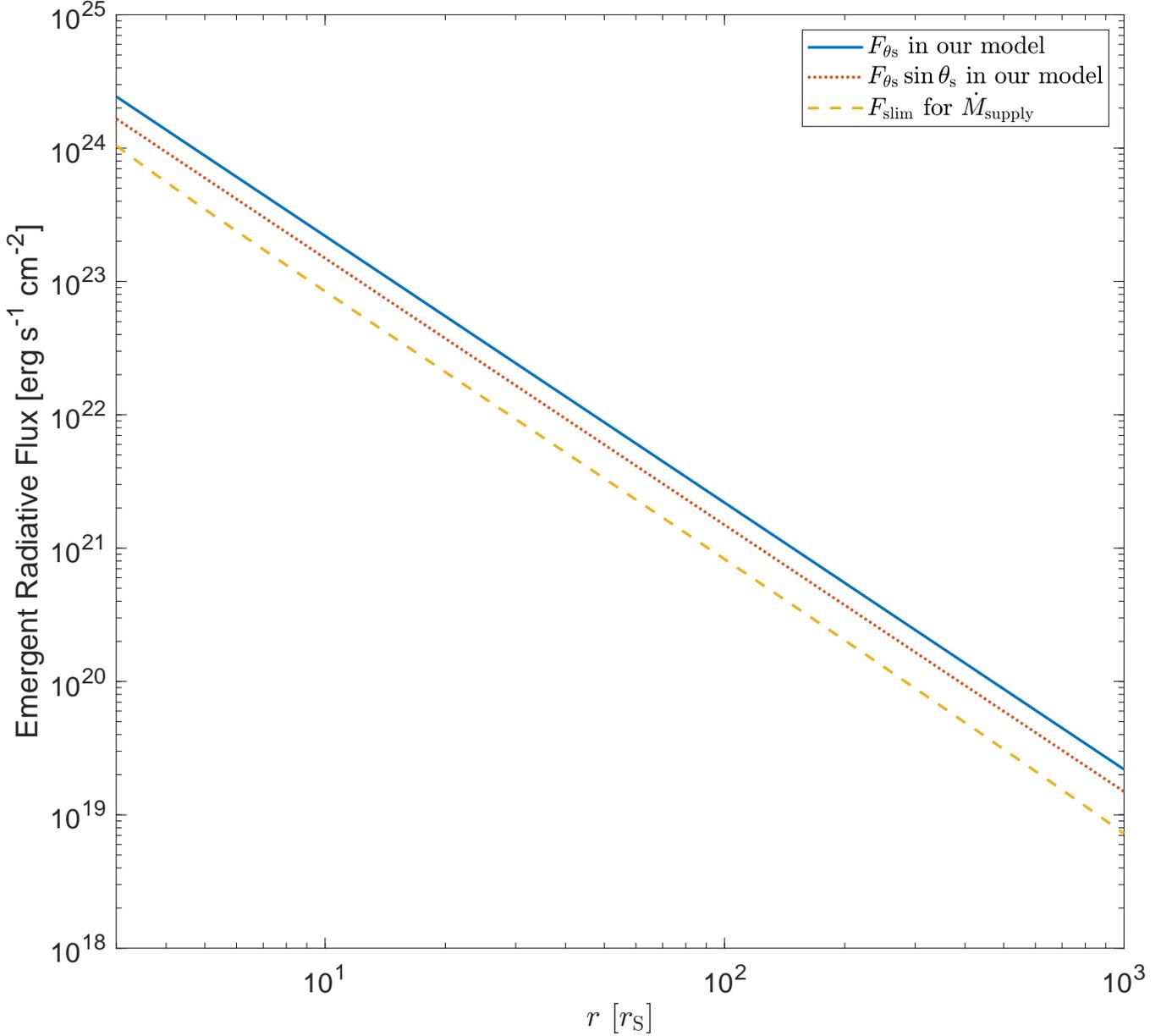}
	\caption{
		Comparison of emergent radiative fluxes. The solid line represents the emergent radiative flux in $\theta$ direction on the surface of the accretion flow in our model, $F_{\theta \mathrm{s}}$, for the solution presented in Figure~\ref{fig_sol1}, the dotted line represents $F_{\theta \mathrm{s}}\sin{\theta_\mathrm{s}}$, with a factor $\sin{\theta_\mathrm{s}}$ to keep the luminosity formula the same as models in cylindrical coordinates such as the slim disk, and the dashed line represents the emergent radiative flux of the slim disk model for the mass supply rate from the environment, $\dot{M}_\mathrm{supply}$.}
	\label{fig_flux}
\end{figure}

The comparison of emergent radiative fluxes are shown in Figure~\ref{fig_flux}. The solid line represents $F_{\theta \mathrm{s}}$, the dotted line represents $F_{\theta \mathrm{s}}\sin{\theta_\mathrm{s}}$, with a factor $\sin{\theta_\mathrm{s}}$ to keep the luminosity formula the same as models in cylindrical coordinates [Equations (\ref{eq_L1}) and (\ref{eq_Ltheta})], and the dashed line represents the emergent radiative flux of the slim disk for the mass supply rate from the environment, $\dot{M}_\mathrm{supply}=6800 L_\mathrm{Edd}/c^2$, calculated with the code for traditional slim disk model in \cite{Jiao09} (not the revised slim disk model whose accretion rate has an upper limit beyond which outflow will be inevitably developed), where we set $\kappa_\mathrm{es}=0.4$ cm$^2$ g$^{-1}$ in accordance with the low metallicity case in this paper, instead of $\kappa_\mathrm{es}=0.34$ cm$^2$ g$^{-1}$ in \cite{Jiao09} corresponding to solar elemental abundances. Note that although we plot the lines all the way to the ISCO at $r=3r_\mathrm{S}$, the self-similar assumptions are generally no longer reliable close to the BH.
We can see that both $F_{\theta \mathrm{s}}$ and $F_{\theta \mathrm{s}}\sin{\theta_\mathrm{s}}$ are significantly larger than $F_\mathrm{slim}$, indicating that the radiative efficiency of our model is larger than that of the slim disk model, even when calculated based on the mass supply rate. This result agrees with the numerical simulations of \cite{Jiang2014,Jiang2019}.

As mentioned above, $F_\theta$ keeps increasing as $\theta$ decreases because the local radiative cooling adds into the flux, which can be clearly seen in Equation (\ref{eq_Qrad}). As the non-self-similar region beyond our calculation boundary is an outflow region, the photons will not be carried into the central BH and we expect that the total radiation will continue to increase in this region, so that the radiative flux obtained in our solution is only a lower limit. This further enhances our conclusion that the radiative efficiency of super-Eddington accretion should be larger than that obtained from the slim disk model.

\section{Summary}\label{summary}
Photon trapping is believed to take place in the inner part of super-Eddington accretion, where the photon diffusion timescale becomes longer than the accretion timescale due to very large optical depth. This effect is quantified in one-dimensional height-integrated models, such as the slim disk model, as radial advection of energy, which participates in the energy equation as an important cooling mechanism. However, when multi-dimensional effects, such as outflow and vertical advection, are considered, this understanding change completely. We investigate the problem in this paper, establishing a new 2D solution of super-Eddington accretion with radial self-similarity. Compared with our previous work \citep{JW11}, the equation of radiative flux is added to the model, and the advective factor is now calculated self-consistently instead of being set as a constant input parameter. We also avoid the problem that the net accretion rate becomes 0, which happens in another self-similar solution of super-Eddington accretion flow presented by \cite{Zeraatgari2020}.

A typical solution of our self-similar model consists of an inflow region around the equatorial plane and an outflow region beyond the inflow. Our calculation starts from the equatorial plane with equatorial boundary conditions derived from reflection symmetry (except for the density coefficient which is an input parameter determined eventually by the accretion rate), and stops at a certain inclination angle regardless of the numerical method that we choose. We interpret this calculation boundary as the physical boundary between regions in the accretion flow which can and cannot be described by self-similar assumptions. The non-self-similar region lies between the calculation boundary and the polar axis, and contains outflow blowing out of the surface of the self-similar region, though the exact strucutre there is not solved in this paper.

We have also tested the solution dependence on input parameters [$\alpha$, $M$, $\rho(\pi/2)$], and find that outflow always exists as long as the input parameters are reasonable values under our assumptions. Note that our model itself does not consider the effects of gas pressure and absorption opacity, so that we have to rely on other theoretical studies or numerical simulations to ensure that neglecting these two effects are feasible. According to the discussion in Section \ref{eqs}, a simple way to ensure that these assumptions are justified is to ensure that the flow is super-Eddington. For an application of our model to observations, we can evalute the applicability based on the Eddington ratio of the observed source. For theoretical studies with a given set of input parameters, we can calculate the effective accretion rate of our model ($\dot{M}_{\mathrm{eff}} \propto r$ in the self-similar region), and compare its lower-limit value calculated at $r_\mathrm{ISCO}=3r_\mathrm{S}$ (the real accretion rate is likely much larger than this lower-limit value because the self-similar region is generally away from the inner boundary) with the Eddington accretion rate. With fixed values of $\alpha$ and $M$, reducing $\rho(\pi/2)$ causes $\dot{M}_{\mathrm{eff}}$ to decrease, and the sizes of the inflow and the outflow regions and the mean advective factor will all decrease accordingly. With fixed values of $\alpha$ and $\rho(\pi/2)$, reducing $M$ causes the normalized accretion rate $\dot{m}=\dot{M}_{\mathrm{eff}} /\dot{M}_\mathrm{Edd}$ to decrease [because $\rho(\pi/2)$ is in cgs units, see Section \ref{f_mean} for a detailed derivation], and the sizes of the inflow and the outflow regions and the mean advective factor will all decrease as well.

The outflow-driving mechanism in our solution is explored based on force analyses in $r$ and $\theta$ directions. We find that in $r$ direction, the radiative force plays an important role in driving the outflow, though the dominant force that resists the gravity is still the centrifugal force. This agrees with numerical simulations of super-Eddington accretion that have presented the force analysis in $r$ direction. Some simulations have found that the radiative force could become the dominant force in $r$ direction around the polar axis, but the structure of accretion flow in that region is likely non-self-similar, and our self-similar solution cannot reach that region. In $\theta$ direction, the radiative force always points away from the equatorial plane as radiation tries to escape the accretion flow, while the $\theta$ component of the centrifugal force acts as an opposing factor, except for $\theta=\pi/2$ on the equatorial plane where the $\theta$ components of all forces are 0 due to reflection symmetry. The total force in $\theta$ direction always points away from the equatorial plane (when $\theta \neq \pi/2$), which drives the material upward towards the polar axis.
Note that in some papers, the material that escapes vertically is also considered outflow, and the outflow-driving mechanism is discussed based on the analysis of vertical ($z$) components of forces in cylindrical coordinates. If we make a similar discussion in $\theta$ direction, we can also say that our flow is radiation-driven, because the radiative force in $\theta$ direction is the dominant force that drives the flow away from the equatorial plane.

The energy conservation of our solution is examined. We find that radial advection is actually a heating mechanism in the inflow due to compression, and a cooling mechanism in the outflow due to expansion. While radiation helps to cool the inflow, it alone cannot maintain the energy balance, and the entropy from compression and viscous heating that is not cooled by radiation is then carried upward by vertical advection. Part of this entropy is released as radiation at higher latitudes where photons can escape more easily, and the rest of the entropy is carried away by the outflow. These effects are represented by the profile of the vertical advection in the energy equation (Figure~\ref{fig_qf}), which is the main cooling mechanism in a large part of the inflow and becomes the main heating mechanism close to the surface, which supplies the enhanced radiation there with accumulated entropy from the inflow and lower part of the outflow.

We also calculate the mean advective factor of our solution, which is 0.29 for the solution presented in Figure \ref{fig_sol1}, smaller than that expected by the slim disk model ($>0.5$ for super-Eddington accretion). This means that more energy is cooled by radiation in our solution than in the slim disk model, which is not surprising because vertical motion in our
solution carries photons towards the surface of the accretion flow, which effectively reduces the photon diffusion time and consequently inhibits the photon trapping effect.
However, for the same mass supply rate, the existence of outflow reduces the net accretion rate, and consequently less viscous heating is generated, which weakens the luminosity. To investigate the eventual effect, we present a comparison of the emergent radiative fluxes of our model and the slim disk model based on the same mass supply rate, so that reduction in the net accretion rate of our model is also accounted for. The emergent radiative flux of our model is still larger than that of the slim disk model, so that the radiative efficiency of our model should still be larger. This result agrees with the numerical simulations of \cite{Jiang2014,Jiang2019}.

The angular momentum conservation of our solution is also analysed. While the viscous torque transfers angular momentum outward and tends to reduce the angular momentum for a fixed region in the rest frame, the advetion of angular momentum due to flow motion compensates this loss and maintains the angular momentum conservation in the steady solution. The radial advection is an angular momentum gain in the inflow, but becomes an angular momentum loss in the outflow, where the balance is upheld solely by the angular momentum gain provided by the vertical advection. This means that the angular momentum per unit mass should decrease in the direction of $v_\theta$, which is in the $\theta$-decreasing direction. However, the value of $v_\phi$ still tends to increase as $\theta$ decreases for a fixed radius, because the distance to the axis of rotation decreases faster (due to the geometry of spherical coordinates) than the angular momentum.

There are some caveats in this work that should be mentioned. First, we only consider the $r\phi$-component of the viscous stress tensor. The inclusion of other components of the viscous stress does not change the formula of advective energy transport, and only changes the formula of the viscous heating, which would still be a heating term anyway, so the general results in this paper should still hold. Secondly, we focus on the extremely radiation-pressure dominated region of the flow and ignored the free-free and bound-free absorption. While this could be justified in the high-temperature region of super-Eddington accretion, these treatments definitely limit the applicability of the model, and should be improved in our future work. Thirdly, numerical simulations have shown that the magnetorotational instability (MRI) is at least an important mechanism in the outward transfer of angular momentum (see its comparison with Reynolds stress in \citealp{Jiang2014,Jiang2019}), so the consideration of magnetic field in the model is also planned in our future work. 
Fourthly, other means of energy transport, e.g. magnetic buoyancy, convection, may be important in a realistic turbulent accretion flow and is not considered in this work, which will be addressed in future.
Lastly, due to restrictions of self-similarity, a steady self-similar inflow-outflow solution requires either $\dot{M}_\mathrm{total}=0$ (as in \citealp{Zeraatgari2020}) or the existence of a non-self-similar region (as in our solution). As the super-Eddington luminosities observed in many sources, such as ULXs, require a rather large total accretion rate of matter whose gravitational energy powers the radiation, we think that our solution is more applicable to observations. The trade-off is that our solution cannot describe the non-self-similar region, so that the emergent radiative flux does not include the contribution of this region and should be considered a lower limit. We are also planning to abandon the self-similar assumptions and establish 2D global solutions of super-Eddington accretion, similar to the 2D global solution of ADAFs in \cite{KG2018}, in our future work.

\begin{acknowledgments}
	We thank the referee for helpful comments. This work is supported by the Natural Science Foundation of China (grant 11703083).
\end{acknowledgments}

\bibliography{refs}{}
\bibliographystyle{aasjournal}

\end{CJK*}
\end{document}